# Inter- to Intra-Layer Resistivity Anisotropy of NdFeAs(O,H) with Various Hydrogen Concentrations

M. Chen[1], K. Iida[1,3], K. Kondo[1], J. Hänisch[2], T. Hatano[1,3], H. Ikuta[1]

[1] Department of Materials Physics, Nagoya University, Nagoya 464-8603, Japan

[2] Institute for Technical Physics, Karlsruhe Institute of Technology, Hermann-von-Helmholtz-Platz 1, 76344 Eggenstein-Leopoldshafen, Germany

[3] JST CREST, Kawaguchi, Saitama 332-0012, Japan

**Abstract**

With molecular beam epitaxy and topotactic chemical reaction, we prepared NdFeAs(O,H) epitaxial thin films with various hydrogen concentrations on 5° vicinal-cut MgO substrates. By measuring the resistivities along the longitudinal and transversal directions, the *ab*-plane and the *c*-axis resistivities ($\rho_{ab}$ and $\rho_c$) were obtained. The resistivity anisotropy $\gamma_\rho \equiv \rho_c/\rho_{ab}$ of NdFeAs(O,H) with various hydrogen concentrations was compared with that of NdFeAs(O,F). At the H concentrations which led to superconducting transition temperatures $T_c$ over 40 K, $\gamma_\rho$ recorded ~100-150 at 50 K. On the other hand, a low $\gamma_\rho$ value of 9 was observed with the mostly doped sample. The exponent $\beta$ of the *ab*-plane resistivity obtained by fitting a power law expression $\rho_{ab}(T) = \rho_0 + AT^\beta$ to the data was close to unity down to low temperature in the vicinity where the second antiferromagnetic phase locates, which may be related to the quantum critical point discussed at the over-doped side of the phase diagram.

1. **Introduction**

   Mixed-anion compounds have recently been attracting considerable attention, since they are rich in emergent phenomena, e.g. acid nitride photocatalysts [1] and acid hydride proton conductors [2]. Iron-based mixed-anion compound *Ln*FeAsO (*Ln* = Nd and Sm) shows superconductivity with a maximum transition temperature $T_c$ of 56 K when oxygen is partially substituted by fluorine (*Ln*FeAsO$_{1-x}$F$_x$) or hydrogen (*Ln*FeAsO$_{1-x}$H$_x$) [3][4][5]. These substitutions lead to electron doping (O$^{2-}$ → H$^-$/F$^-$ + e$^-$), therefore, the carrier concentration is determined by the amount of F or H. The decisive difference between fluorine and hydrogen is the solubility limit: the latter can be substituted up to $x \sim 0.8$ [6], whereas the former about $x = 0.2$ at ambient pressure [7]. *Ln*FeAsO crystallizes in a laminated structure with alternating *Ln*O charge reservoir and FeAs conduction layers along the *c*-axis direction. Therefore, the electromagnetic properties of *Ln*FeAsO are anisotropic. From the viewpoint of application, however, a lower electromagnetic anisotropy is favorable. Compared with F-doped *Ln*FeAsO, hydrogen substituted *Ln*FeAsO has a more three-dimensional like Fermi surface, and thus a lower anisotropy can be expected [8]. Additionally, the character of the Fermi surface may depend on the carrier doping level as was shown for another Fe-based superconductor, BaFe$_2$As$_2$ [9][10]. Hence, it is interesting to explore how the electromagnetic anisotropy of *Ln*FeAsO$_{1-x}$H$_x$ changes as a function of doping level.

   The resistivity anisotropy $\gamma_\rho$ defined as the ratio $\rho_c/\rho_{ab}$ where $\rho_c$ and $\rho_{ab}$ are the out-of-plane and in-plane resistivities, is related to the mass anisotropy $\gamma_m \equiv m_c/m_{ab}$ where $m_c$ and $m_{ab}$ are the out-of-plane and in-plane effective masses, respectively. Polycrystalline *Ln*FeAsO samples with different hydrogen contents have been used for exploring physical quantities [11], but they are not ideal platforms for investigating the electromagnetic anisotropy. Unfortunately though, fabrication of large enough H-doped *Ln*FeAsO single crystals seems to be still difficult, and only a few studies have been reported measuring the resistivity anisotropy of doped $Ln$FeAsO[12][13][14][15]. Recently, SmFeAsO$_{0.9}$H$_{0.1}$ single crystals have been grown by combining a Na$_3$As flux and a high-pressure synthesis method [15]. The $\gamma_\rho$ of the resultant crystals increased from ~3 at 300 K to 8 at 50 K. However, it is not clear how the resistivity anisotropy depends on the hydrogen content, in particular, at high concentration.

Recently, hydrogen doped $Ln$FeAsO epitaxial thin films have been fabricated via a topotactic chemical reaction [16][17][18]. In one of these works, we showed that hydrogen contents can be varied by controlling the processing temperature $T_{\text{proc}}$ [18], which enables us to investigate the physical quantities as a function of carrier concentration. Additionally, we have successfully grown off-axis NdFeAsO thin films using vicinal substrates in an earlier work [19], from which the out-of-plane resistivity can be extracted. Combining these two methods, NdFeAs(O,H) thin films having different H concentrations were grown on vicinal-cut MgO substrates in the present study. We measured the transport properties of these films and discussed how the physical quantities and their anisotropy depend on the carrier concentration.

2. **Experimental**

The growth of parent NdFeAsO thin films with thicknesses of 20-30 nm was carried out by molecular beam epitaxy (MBE) at 800 °C. Fe, As, NdF$_3$, Fe$_2$O$_3$ and Ga were charged in Knudsen cells. Here, oxygen was supplied via the thermal decomposition of Fe$_2$O$_3$, whereas Ga was used to remove F according to NdF$_3$+Ga→Nd+GaF$_3$. The growth procedure was described in detail in Ref. [20]. In-situ reflection high-energy electron diffraction (RHEED) was utilized for monitoring the film growth. Instead of normal MgO(001) substrates, 5×10 mm$^2$ sized vicinal-cut substrates with the [001] axis inclined by ~5° with respect to the direction normal to the surface were used for extracting the out-of-plane resistivity [fig. 1(a)]. After the film growth, X-ray diffraction (XRD) analysis was performed: $2\theta/\omega$ scans were used for evaluating the phase purity and the $c$-axis lattice parameter. Because the film grows with an offset angle on a vicinal-cut substrate, we also conducted $\omega$-scans with the angle $2\theta$ fixed to the 003 reflection of NdFeAsO to determine the offset angle $\alpha$ precisely. The angle $\omega$ that maximizes the diffraction intensity corresponds to $\alpha$.

Doping hydrogen into NdFeAsO was conducted similarly to Ref. [18]. As-grown films were cut into smaller samples having a $5 \times 5$ mm$^2$ size. The sample was placed in a Al$_2$O$_3$ crucible filled with CaH$_2$ powder (~0.8 g) as a hydrogen source. The whole arrangement was placed in a quartz tube (outer diameter 12 mm and length 150 mm) and then sealed under vacuum. Importantly, the surface of the film had to be physically in contact with the CaH$_2$ powder. Then the quartz tube was heated to $T_{\text{proc}}$ = 440~500 °C where it was held for 36 hours. The doping level of H was controlled by adjusting the processing temperature.

The H-doped NdFeAsO films were photolithographically patterned and Ar-ion-beam etched to make two kinds of bridges (50 μm width and 0.5 mm long), as shown in fig. 1(b). They are used for measuring the transport properties in the longitudinal direction (abbreviated as L-bridge) and in the transverse direction (T-bridge), respectively [fig. 1(a)]. A four-probe method was employed with a bias current of 1 μA to measure resistivity using a physical properties measurement system (PPMS). The onset superconducting transition temperature $T_c^{\text{onset}}$ was determined as the intersection between the fit to the normal state resistivity and the steepest slope of the transition region. The zero resistivity temperature $T_c^0$ was defined by a resistivity criterion, $\rho_{\text{cri}} = 0.05$ μΩcm, which corresponds to the noise level of our experimental set-up (see Supplementary information). In the longitudinal direction, the current flows only in the *ab*-plane, so the longitudinal resistivity $\rho_\text{L}$ is the same as the *ab*-plane resistivity (i.e. $\rho_{ab} = \rho_\text{L}$), whereas in the transverse direction, the current flows both in the *ab*-plane and along the *c*-axis. Therefore, $\rho_c$ can be calculated from the transverse resistivity ($\rho_\text{T}$) by the following equation [21]

$$\rho_c = (\rho_\text{T} - \rho_\text{L}\cos^2\alpha)/\sin^2\alpha. \quad (1)$$

The Hall resistivity $\rho_{xy}$ was measured in the field range of up to $\mu_0 H = \pm 9$ T at 50 K. As expected, the slope corresponding to the Hall coefficient $R_\text{H}$ was always negative, indicating that the dominant carrier is electron. The carrier density *n* was evaluated using the single carrier model formula

$$R_\text{H} = -\frac{1}{ne}. \quad (2)$$

Here, *e* is the elementary charge. Note that for the Hall effect measurements, the bias current was supplied along the longitudinal direction, whereas the Hall voltage was measured along the transverse direction, which is the tilted direction. This may lead to a deviation between the actual carrier density and the measured one [22]. The upper critical field $H_{c2}$ was determined by measuring the field dependence of resistivity and using a criterion of 90% of the normal state resistivity.

## 3. Results and discussion

### 3.1 Structural characterization

The specifications of the samples studied here are listed in table 1. For convenience, these samples are named in the order of their carrier densities. We used five as-grown films. Samples B and E were prepared from the same as-grown film and all other samples were prepared from different as-grown films. As stated above, the doping level of H can be controlled by adjusting the processing temperature $T_\text{proc}$. However, this applies only to the samples derived from the same as-grown film. That is why the correlation between *n* and $T_\text{proc}$ seems to be weak (see table 1).

XRD patterns of samples B and E, both originating from the same as-grown film, are compared in fig. 2. For comparison, the pattern of the as-grown film is also shown. Although small peaks arising from impurities were observed for sample B, the XRD patterns of all samples including those not shown here dominantly consist of the 00*l* peaks of the NdFeAs(O,H) phase and the substrate peaks, indicating that they are *c*-axis oriented. The intensity of the 002 peaks of the doped samples in fig. 2 were small and almost indistinguishable from the background. Our study using an ordinary (not a vicinal cut) substrate shows that the intensity of the 002 peak relative to the other peaks decreases with H doping (see Supplementary information). Hence, the weaker intensity is, at least partially, due to the slight change in the crystal coordinates and the difference in the atomic scattering factors of oxygen and hydrogen. However, a slight degradation of the crystallinity cannot be ruled out completely. By comparing the doped samples with the as-grown one, clear shifts of the 00*l* peaks to higher angles are recognized, which reflects the reduction in the *c*-axis length. The evaluated *c*-axis lengths for all samples are summarized in table 1. The carrier density and the *c*-axis length are well correlated, which is consistent with our previous results on NdFeAs(O,H) grown on ordinary MgO(001) substrates [18] and found in bulk H-doped or F-doped *Ln*FeAsO samples [23][24].

### 3.2 Resistivity measurements at zero magnetic field

The temperature dependencies of the resistivity in the transversal and longitudinal directions for all samples are summarized in fig. 3(a). As can be seen, the transversal resistivity is always larger than the longitudinal one because $\rho_T$ contains the $\rho_c$ component. For all samples, each pair of transversal and longitudinal bridges fabricated from the same film possess the same $T_c^{onset}$ and $T_c^0$, proving that the films are chemically homogeneous.

The carrier density dependencies of both $T_c^{onset}$ and $T_c^0$ are plotted in fig. 3(b). Both $T_c^{onset}$ and $T_c^0$ decreased with increasing carrier concentration. As the behavior of $T_c^{onset}$ and $T_c^0$ on $n$ are similar, $T_c^{onset}$ is regarded as $T_c$ from here on for simplicity.

The temperature dependencies of the in-plane resistivity $\rho_{ab}(T)$ and the out-of-plane resistivity $\rho_c(T)$ of all samples are plotted in figs. 4(a) and (b), respectively. All $\rho_{ab}(T)$ curves exhibited a metallic behavior irrespective of $n$. However, the shape of the $\rho_{ab}(T)$ curve for samples E and F is obviously different from that of samples A, B and C. The $\rho_{ab}(T)$ curves of the former samples have a positive curvature, while those of the latter samples are slightly negative. On the other hand, sample E shows an almost linear temperature dependence from $T_c$ to 120 K (see Supplementary information). This will be discussed later. The behavior of $\rho_c(T)$ shown in fig. 4(b) is converse to that of $\rho_{ab}(T)$. $\rho_c$ depends almost linearly to temperature at larger $n$, while the $\rho_c(T)$ curves of the less doped samples have a negative curvature. Moreover, the temperature dependence of $\rho_c$ for samples A-C showed a peak at around 250-275 K, which may correspond to a crossover from incoherent conduction at high temperatures to coherent conduction at low temperatures as was observed, for instance, also in the BaFe2As2 system [25]. Note that the correlation length $L$ evaluated by the Williamson-Hall analysis is close to the film thickness (20~30 nm) and independent of carrier concentration (see Supplementary information). Thus, the variation of $\rho_c$ is not due to the change in the microstructure but the H content.

Figure 4(c) shows the $n$ dependence of $\rho_{ab}$ at 300 and 50 K. $\rho_{ab}$ at both 300 and 50 K decrease with increasing $n$ up to $4 \times 10^{21}$ /cm$^3$. An abrupt rise in $\rho_{ab}$ at around $n = 4 \times 10^{21}$ /cm$^3$ is observed at both 300 and 50 K. This carrier concentration coincides with that where a quantum critical point (QCP) may be present as will be discussed later. Unlike $\rho_{ab}$, $\rho_c$ at 300 and 50 K decrease monotonically with increasing $n$, as shown in fig. 4(d).

The resistivity anisotropy $\gamma_\rho$ is shown in fig. 5(a) as a function of temperature. For samples A-E, $\gamma_\rho$ increases monotonically with decreasing $T$. This behavior is the same as that of optimally F-doped NdFeAsO thin films [19]. The maximum $\gamma_\rho$ of the samples with a $T_c$ above 40 K reaches about 100 to 150 just above $T_c$. The temperature dependence of $\gamma_\rho$ becomes weaker and the value of $\gamma_\rho$ decreases as $n$ increases. Finally, $\gamma_\rho$ of the mostly doped sample has a different temperature dependence. It decreases with decreasing temperature and records a minimum of about 8 around 100 K before increasing at lower temperatures as shown in fig. 5(b). Although the variation in the temperature dependence of the present sample F is small, the minimum of $\gamma_\rho$ might be an indication of a phase transition at around 100 K. In fact a structural transition was observed for SmFeAs(O,D) and LaFeAs(O,H) at about 100 K in the over-doped region, where the $c$-axis length showed a minimum [6][26]. As a shorter $c$-axis length means an increased connectivity between the FeAs conducting layers, this may explain the minimum of the resistivity anisotropy.

Figure 5(c) shows the carrier density dependence of $\gamma_\rho$ at 50 K for H-doped NdFeAsO. The films with higher $T_c$ tend to have a larger anisotropy. The peak of $\gamma_\rho$ is located at around $n = 2.2 \times 10^{21}$ /cm$^3$, which corresponds to the high-$T_c$ region [see fig. 3(b)]. It is notable that the optimally F-doped sample has a larger anisotropy than the H-doped samples [19]. This might be explained by the more two-dimensional character of the Fermi surface of F-doped samples [8]. Alternatively, the high reactivity of fluorine may have introduced some damage to the crystal during the fluorination that disturbed the interplane conductivity.

3.3 In-field resistivity measurements of samples C and E

The upper critical fields $H_{c2}$ of sample E determined from the in-field $\rho_{ab}$ measurements for both $H \parallel c$ and $\parallel ab$ are summarized in fig. 6(a). The respective slopes of the temperature dependence are -4.8 T/K and -14.8 T/K near $T_c$ for $H \parallel c$ and $\parallel ab$, which gives an anisotropy of 3.1. For single-band superconductors, the anisotropy of the upper critical field $\gamma_{H_{c2}} (= H_{c2\perp}/H_{c2\parallel}$, $H_{c2\perp}$ and $H_{c2\parallel}$ are the upper critical fields for $H \parallel ab$ and $\parallel c$) equals to $\sqrt{\gamma_\rho} (= \sqrt{\gamma_m})$. This relation does not necessary hold for multi-band superconductors. Nevertheless, the value of $\gamma_{H_{c2}}$ is relatively close to the square root of the resistivity anisotropy [$\gamma_\rho(50\ \text{K}) \sim 15$] of sample E.

We also measured the resistivity under magnetic fields in the flux-flow regime of sample E, and found that the angle dependence of resistivity under various magnetic fields collapses to a single curve when plotted as a function of an effective magnetic field $\mu_0 H_{\text{eff}} = \mu_0 H / f_T(\theta)$, where $f_T(\theta)$ is the Tinkham formula given by [27]

$$f_T(\theta) = \frac{1}{2\gamma_{H_{c2}}^2 \sin^2\theta} \left[ \sqrt{\cos^2\theta + 4\gamma_{H_{c2}}^2 \sin^2\theta} - |\cos\theta| \right]. \tag{3}$$

Here, $\theta$ is the angle between the magnetic field and $c$-axis. Figure 7 shows the data taken at 26.5 K. The anisotropy parameter $\gamma_{H_{c2}}$ was determined as 3.2±0.1 from the scaling of fig. 7(b), consistent with the value obtained from the temperature slope of $H_{c2}$. It is also noticeable that the dip in the resistance curve is shifted from 270° to a higher angle at low magnetic fields due to the tilted growth of the film and the corresponding non-parallelity of $B$ and $H$ (see fig. 7(c)) [28]. The shift of the resistance dip explains the slight deviation of the data near zero effective field from the scaling curve in fig. 7(b). Resistivity scaling at slightly higher temperatures (27.0 and 27.5 K) with the same value of $\gamma_{H_{c2}}$ were of similar quality. On the assumption that the relaxation time of carriers is isotropic, the resistivity anisotropy is equal to the mass anisotropy. In our previous study, the mass anisotropy was different in the normal and superconducting states for under- and optimally doped NdFeAsO [19]. Hence, the rough match of the mass anisotropy in the normal state with the one in the superconducting state may be one of the distinct features of the over-doped side of the phase diagram. Fig. 6(b) compares $H_{c2}$ of sample E with that of a less doped one (sample C). For $H \parallel ab$, $H_{c2}$ of both samples are almost identical. On the other hand, for $H \parallel c$, sample E shows a higher $H_{c2}$ at the same reduced temperature $T/T_c$. This is probably due to an increased coupling between FeAs layers with increasing doping.

3.4 Exponent of power law behavior $\rho_{ab}(T) = \rho_0 + AT^\beta$

In order to quantify the behavior of the temperature dependence of $\rho_{ab}$, we examined the exponent $\beta$ by assuming $\rho_{ab}(T) = \rho_0 + AT^\beta$ in the temperature range from just above $T_c$ to 120 K, where $\rho_0$ is the residual resistivity at zero kelvin and $A$ is a constant. The obtained $\beta$ is plotted against the carrier density in fig. 8(a), which shows a dome-like shape. A similar behavior was also found for polycrystalline $Ln$FeAs(O,H) [11].

To investigate the exponent $\beta$ in more detail, we employed the formula $\beta(T) = \frac{d\ln(\rho_{ab}-\rho_0)}{d\ln T}$ to extract the temperature dependence of $\beta$ over a wide temperature range [fig. 8(b)], where the $\rho_0$ value obtained by the fitting described above was used. For all samples, $\beta$ is close to 2 at the temperature very close to $T_c$, indicative of a Fermi liquid state. A similar behavior was also found in FeSe$_{0.89}$S$_{0.11}$ under various pressures [29]. At higher temperatures, $\beta$ is smaller than 2. For example,

$\beta$ is almost constant around 1.5 for a wide temperature range from 200 K to close to $T_c$ as shown in fig. 8(b) for sample B, consistent to the result of the fitting mentioned above.

At $n \cong 4.2 \times 10^{21}$ /cm$^3$, on the other hand, $\beta$ was close to 1 over the temperature range from 150 K to $T_c$, and the general trend of the resistivity curve changed from positive ($\beta > 1$) to negative ($\beta < 1$) curvature with further increasing $n$. Intriguingly, a rather abrupt change in the $\rho_{ab}$ value was observed at this carrier density as pointed out above (fig. 4(c)). The $T$-linear behavior of resistivity has often been associated with a quantum critical point at $T = 0$. However, the $T$-linear behavior usually extends to a fan-like region around the quantum critical point on the phase diagram [30], although here it is observed only in a strip-like region. The electronic phase diagram of H-doped LaFeAsO reported by Hiraishi *et al.* shows that the second antiferromagnetic phase (AFM2) and superconductivity may coexist in the over-doped region [26]. The Néel temperature decreases with the application of a high pressure, and the presence of a pressure-induced quantum critical point (QCP) was suggested from NMR measurements [31]. The anomalous behaviors we have observed at $n \cong 4.2 \times 10^{21}$ /cm$^3$ might be a glimmering of this QCP at ambient pressure, although whether they are indeed related or not is an open question for future studies. Note though that the actual carrier density might be slightly different as mentioned in the experimental section.

For larger values of $n$, $\beta$ is less than 1 in the whole temperature range. For a SmFeAsO$_{0.9}$H$_{0.1}$ single crystal, an exponent $\beta = 0.47$ was reported and associated to the spin-frozen state that locates between AFM1 and the superconducting phases [15]. The electronic phase diagram of $Ln$FeAsO shows that AFM1 and AFM2 exist close to the superconducting phase in the under-doped and the over-doped regimes, respectively [6][26]. It might be possible to explain the small $\beta$ value in the over-doped region similarly.

## 4. Conclusion

NdFeAs(O,H) epitaxial thin films with different doping concentrations were grown on vicinal-cut MgO substrates by MBE. With the combination of F-doped samples in the lower carrier density regime, a large part of the phase diagram can be mapped. The temperature dependence of both $\rho_{ab}$ and $\rho_c$ exhibits metallic behavior irrespective of the doping level for all samples studied in this work. The extracted resistivity anisotropy $\gamma_\rho$ showed a strong doping dependence. At the H concentrations which led to superconducting transition temperatures ($T_c$) over 40 K, $\gamma_\rho$ recorded ~100-150 at 50 K. With increasing the carrier concentration, $\gamma_\rho$ eventually reached a remarkably small value around 8. We extracted the $T$-dependent exponent $\beta$ of $\rho_{ab}(T)$ and discussed the presence of a possible QCP at the over-doped side of the phase diagram.

**Acknowledgements**
This work was supported by JST CREST Grant Number JPMJCR18J4, JSPS Grant-in-Aid for Scientific Research (B) Grant Number 20H02681 and Japan-German Research Cooperative Program between JSPS and DAAD, Grant number JPJSBP120203506.

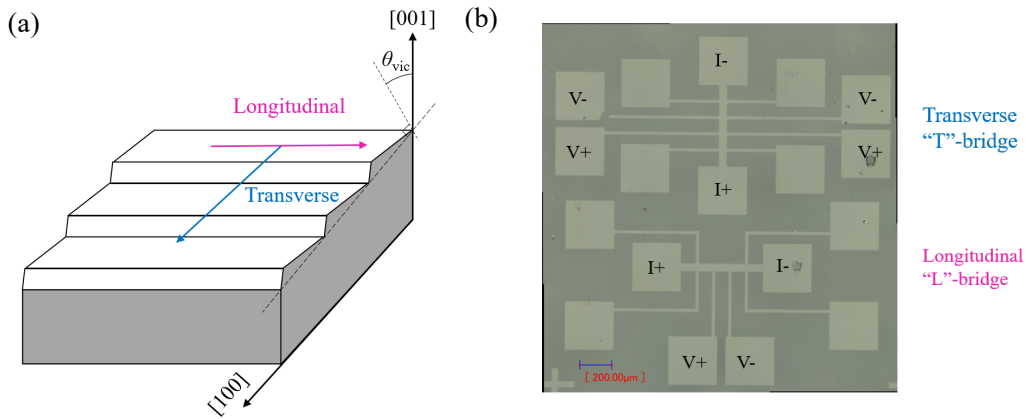

Figure 1. (a) Schematic illustration of a vicinal-cut substrate. The [001] axis is tilted from the direction normal to the substrate surface by $\theta_{vic}$. (b) An example of a top-view optical micrograph of the micro-bridges. A bridge running along the transverse direction is labeled as a T-bridge, whereas a bridge parpendicular to the transverse direction is labeled as an L-bridge.

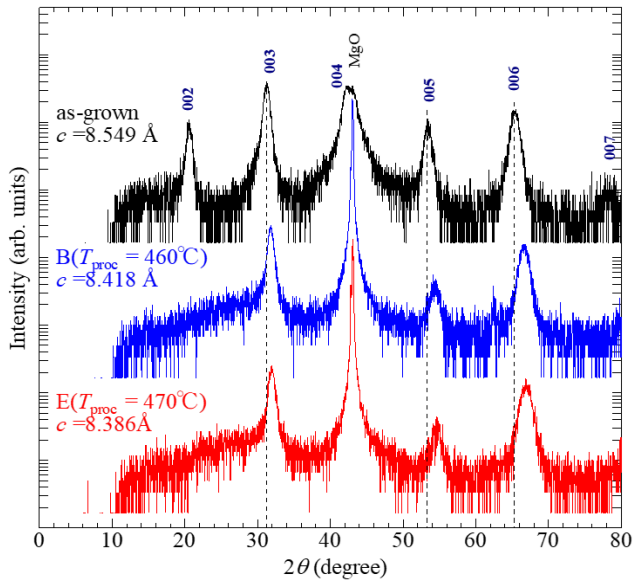

Figure 2. The $2\theta/\omega$ scans of samples B (blue) and E (red), which were processed at 460°C and 470°C, respectively. For comparison, the $2\theta/\omega$ scan of a pristine film (black) with no doping is also shown.

Table 1. Sample name, carrier density $n$, processing temperature $T_{proc}$ of topotactic reaction, the $c$-axis lattice parameter, the growth angle (offset angle of the sample) $\alpha$, and the onset $T_c$ ($T_c^{onset}$) of the samples of this study. The samples are named in the order of the carrier density for convenience.

| Sample name | $n$ ($10^{21}$ /cm$^3$) | $T_{proc}$ (°C) | $c$-axis (Å) | $\alpha$ (°) | $T_c^{onset}$ (K) |
|---|---|---|---|---|---|
| A | 1.51 | 440 | 8.482±0.009 | 5.75 | 50.5 |
| B | 2.28 | 460 | 8.418±0.016 | 5.81 | 42.1 |
| C | 2.75 | 450 | 8.418±0.003 | 4.95 | 43.1 |
| D | 3.99 | 500 | 8.424±0.012 | 6.21 | 37.4 |
| E | 4.23 | 470 | 8.386±0.005 | 5.92 | 27.8 |
| F | 5.03 | 470 | 8.361±0.004 | 5.93 | 24.3 |

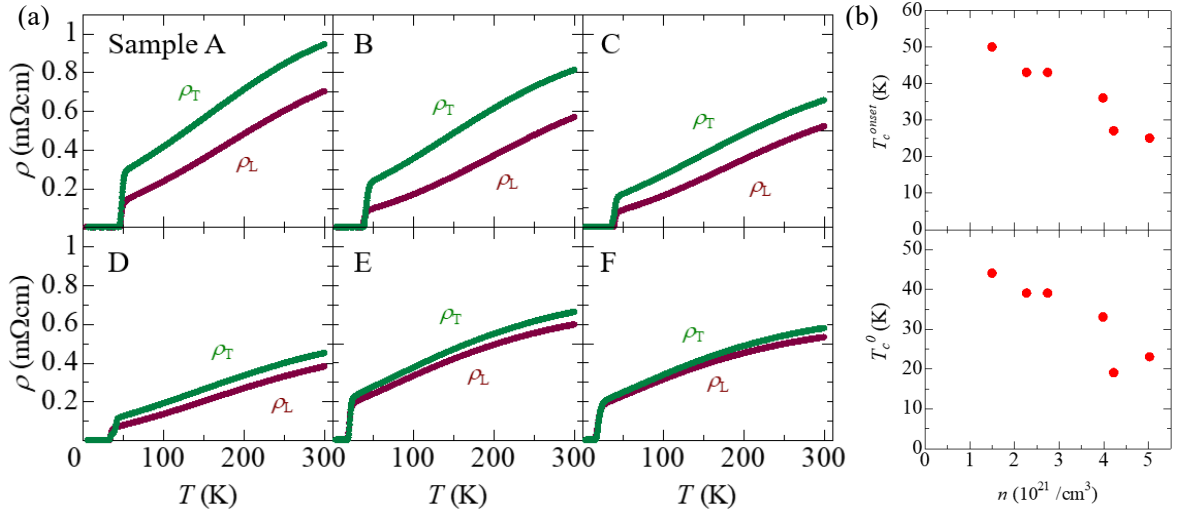

Figure 3. (a) The temperature dependencies of $\rho_L$ (copper) and $\rho_T$ (green) for samples A-F. (b) $T_c^{onset}$ and $T_c^0$ as a function of the carrier density for NdFeAs(O,H).

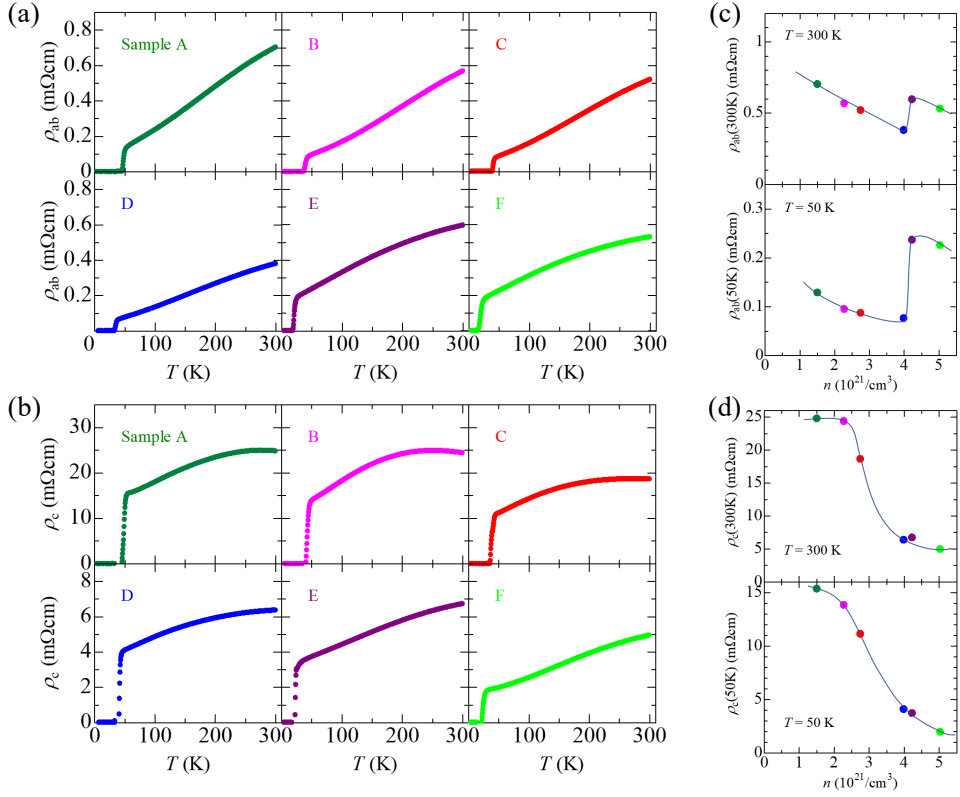

Figure 4. (a-b) The temperature dependence of $\rho_{ab}$ (a) and $\rho_c$ (b) of samples A (green), B (fuchsia), C (red), D (blue), E (copper) and F (lime). (c-d) The carrier density dependencies at 300 and 50 K for (c) $\rho_{ab}$ and (d) $\rho_c$, respectively. For sample A, the resistivities ($\rho_{ab}$ and $\rho_c$) at 50 K were determined by extrapolating the result of linear regression in the temperature range of 55 K < $T$ < 65 K to $T$ = 50 K for its high $T_c^{onset}$ (= 50 K). The solid line is a guide to the eye.

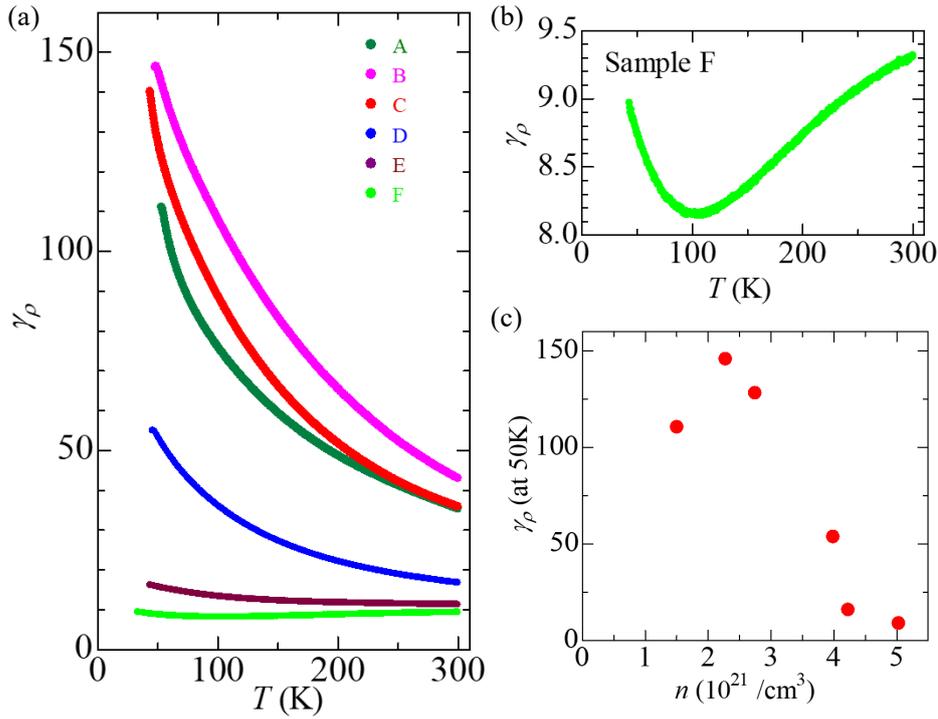

Figure 5. (a) The temperature dependence of the resistivity anisotropy $\gamma_\rho(T) = \rho_c/\rho_{ab}$ of samples A-F. (b) An enlarged view of $\gamma_\rho(T)$ of sample F. (c) The carrier density dependence of $\gamma_\rho(T)$ of NdFeAs(O,H).

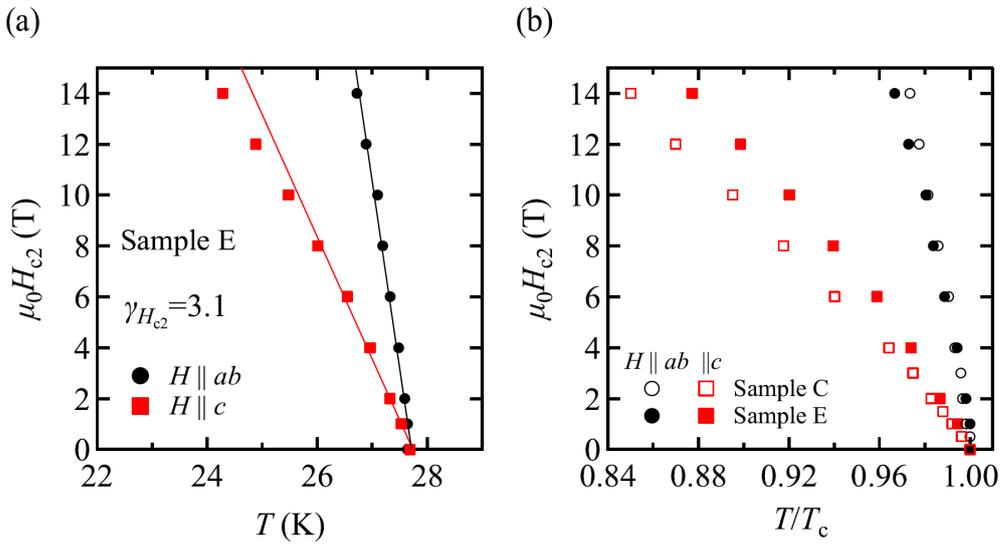

Figure 6. (a) Temperature dependence of $H_{c2}$ of sample E for $H \parallel ab$ and $\parallel c$. The solid lines show the results of linear fitting to the data up to 8 T. (b) Comparison between samples C and E.

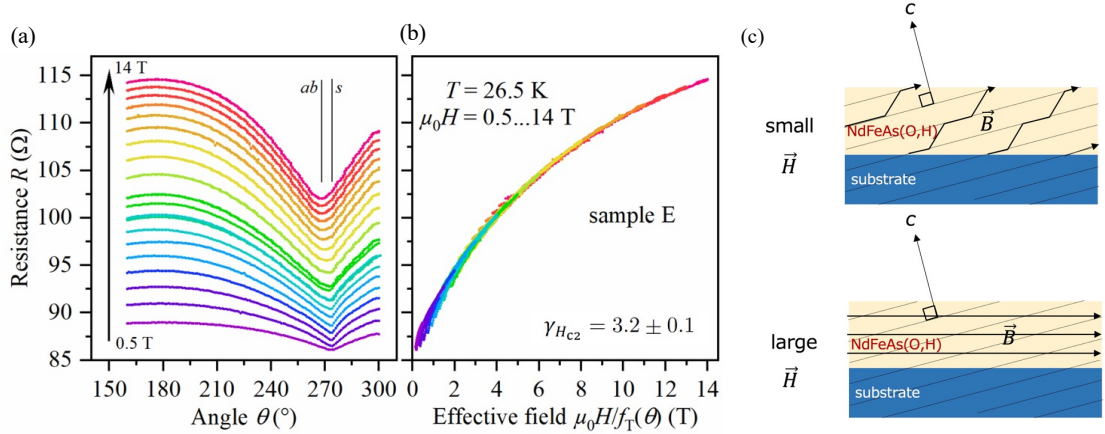

Figure 7. (a) Angular dependence of the resistance at 26.5 K. The magnetic fields were increased by 0.5 T up to 5.0 T and by 1.0 T further on. A shift of the resistance dip from $H \parallel ab$ towards the substrate surface direction (*s*) at low magnetic fields due to non-parallelity of *B* and *H* is clearly observed. (b) 2D (Tinkham) scaling of the data shown in (a). An anisotropy value of $\gamma_{H_{c2}} = 3.2 \pm 0.1$ yielded the best result. (c) Schematic illustrations of how the direction of *B* changes with the strength of the applied magnetic field. When a weak magnetic field is applied parallel to the substrate, *B* is almost parallel to the crystallographic *ab*-plane due to the trapping of *B* along the Nd(O,H) blocking layers. On the other hand, as a strong magnetic field is applied, the direction of *B* is parallel to that of *H*.

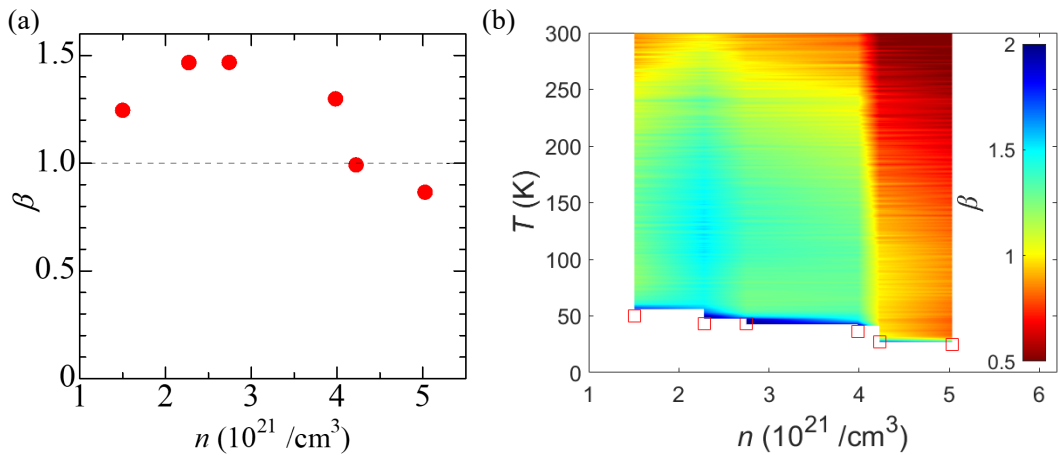

Figure 8. (a) Relationship between the exponent $\beta$ and $n$. The exponent $\beta$ was obtained by fitting the power-law form $\rho(T) = \rho_0 + AT^\beta$ to the resistivity data in the temperature range of $T_c < T \leq 120$ K. The dashed line indicates $\beta = 1$. (b) The carrier density–temperature phase diagram. The colour map represents the exponent $\beta$.

**Supplementary Information on "Inter- to Intra-Layer Resistivity Anisotropy of NdFeAs(O,H) with Various Hydrogen Concentrations"**

M. Chen[1], K. Iida[1,3], K. Kondo[1], J. Hänisch[2], T. Hatano[1,3], H. Ikuta[1]

[1] Department of Materials Physics, Nagoya University, Nagoya 464-8603, Japan

[2] Institute for Technical Physics, Karlsruhe Institute of Technology, Hermann-von-Helmholtz-Platz 1, 76344 Eggenstein-Leopoldshafen, Germany

[3] JST CREST, Kawaguchi, Saitama 332-0012, Japan

1. **Determination of the onset superconducting transition temperature $T_c^{\text{onset}}$ and zero resistance temperature $T_c^0$**

   The onset superconducting transition temperature $T_c^{\text{onset}}$ was defined as the intersection between the fit to the normal state resistivity and the steepest slope of the transition, as shown in Fig. S1(a). Zero resistivity temperature $T_c^0$ was defined by a resistivity criterion, $\rho_{\text{cri}} = 0.05\ \mu\Omega\text{cm}$, which is the noise level of our experimental set-up (Fig. S1(b)).

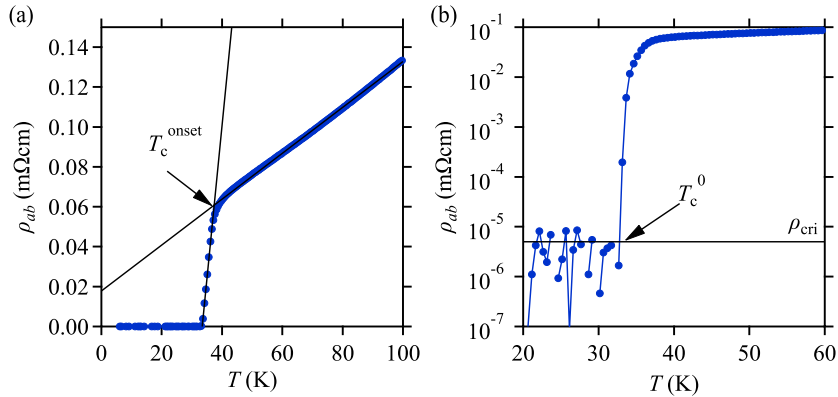

Figure S1 (a) Temperature dependence of $\rho_{ab}$ for sample D. The lines are fits to the resistivity curves in the vicinity of the superconducting transition and the normal state. The onset superconducting transition temperature $T_c^{\text{onset}}$ is defined as the intersection of the two lines. (b) Semi-logarithmic plot of (a) in the vicinity of the superconducting transition.

2. **$2\theta/\omega$ scans of NdFeAs(O,H) films grown on CaF₂ substrates**

   Figure S2(a) shows the $2\theta/\omega$ x-ray pattern of a NdFeAs(O,H) film grown on a CaF₂ substrate. For comparison, the data of the as-grown film are also plotted. Both data are normalized to the 004 peak of the substrate. Obviously, the 002 peak of NdFeAs(O,H) is much weaker than that of NdFeAsO.

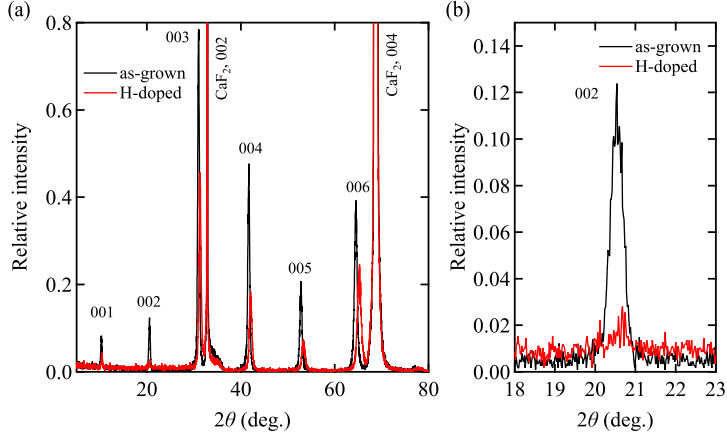

Figure S2 (a) The $2\theta/\omega$ scans of NdFeAs(O,H) grown on CaF$_2$ substrates. (b) An enlarged view around the 002 reflection. The data were normalized to the 004 peak of the CaF$_2$ substrates.

## 3. Resistivity curves $\rho_{ab}(T)$

Figure S3 shows the temperature dependence of the in-plane resistivity for all samples studied in this work. The normal state resistivity from $T_c$ to 120 K was fitted with $\rho_{ab}(T) = \rho_0 + AT^\beta$ as indicated by the dashed lines, and the temperature exponents β are shown in the right of the figures.

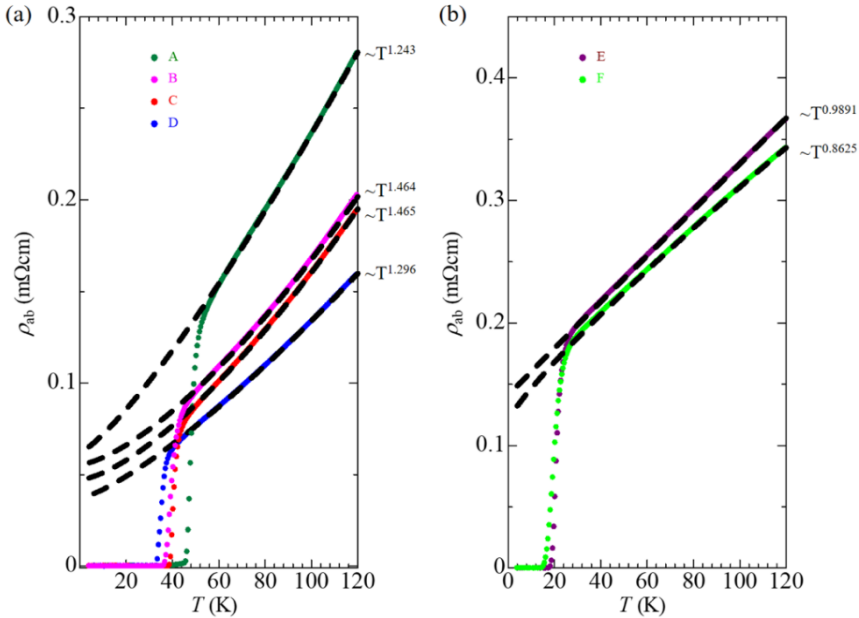

Figure S3 (a) Temperature dependence of $\rho_{ab}$ for samples A-D and (b) samples E-F in the temperature range from $T_c$ to 120 K. As can be seen, the exponent β is close to unity for sample E.

## 4. Williamson-Hall plot for analyzing the correlation length

Figure S4(a) shows the Williamson-Hall plots for all samples (A-E). The correlation length $L$ can be evaluated by the following formula:

$$L = 0.9/2y,$$

where $y$ is the $\beta_{2\theta/\omega}\sin\theta/\lambda$-intersection [S1]. The evaluated $L$ values as a function of carrier concentration is shown in Fig. S4(b). As can be seen, $L$ is close to the film thickness and independent of carrier concentration. Thus, the variation of $\rho_c$ is not due to the change in the microstructure but the H content.

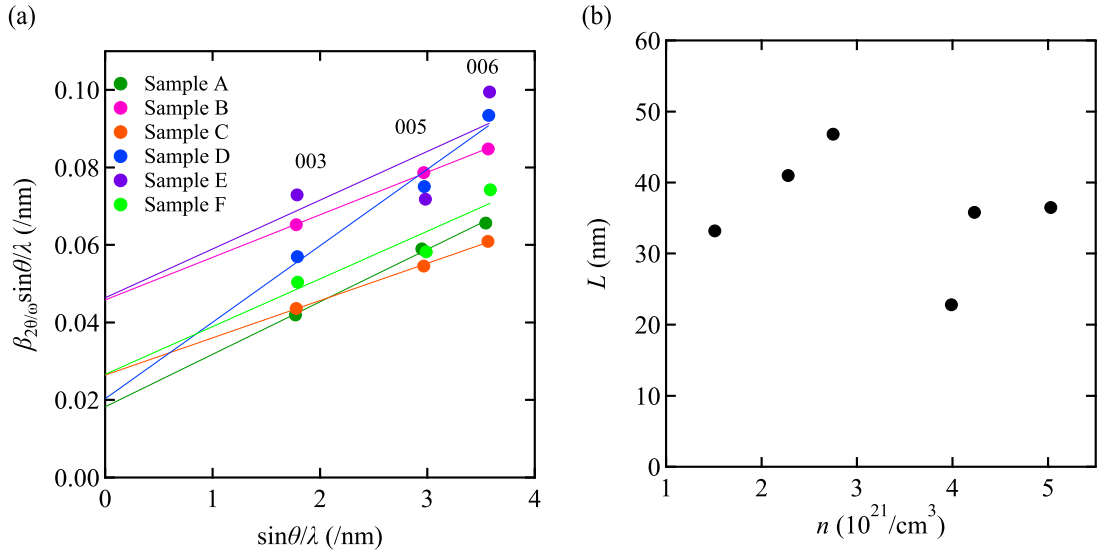

Figure S4 (a) Williamson-Hall plots for all samples (A-E). (b) The correlation length $L$ as a function of carrier concentration.

Reference

[S1] H-M. Wang, J-P. Zhang, C-Q. Chen, Q. Fareed, J-W. Yang, M. Asif Khan, *Appl. Phys. Lett.* **81**, 604 (2002).